\documentstyle[preprint,prb,aps]{revtex}

\tighten

\begin{document}

\title{Equilibrium properties of a Josephson junction
			ladder with screening effects.}

\author{Juan J.\ Mazo $^{(a),(b)}$ and Jos\'e C. Ciria $^{(b)}$}

\address{$^{(a)}$ Departamento de F\'{\i}sica de la Materia 
Condensada and Instituto de Ciencia de Materiales de Arag\'on.
Consejo Superior de Investigaciones Cient\'{\i}ficas.\\
Universidad de Zaragoza, 50009 Zaragoza, Spain}
\address {$^{(b)}$ Dipartimento di Fisica. Universit\`{a} di Roma
Tor Vergata\\ Via della Ricerca Scientifica 1, 00133 Roma, Italy.}

%\date{\today}
%\date{February 26, 1995}

\maketitle

\begin{abstract}
In this paper we calculate the ground state phase diagram of a Josephson
Junction ladder when screening field effects are taken into account. We
study the ground state configuration as a function of the external field,
the penetration depth and the anisotropy of the ladder, using different
approximations to the calculation of the induced fields. A series of
tongues, characterized by the vortex density $\omega$, is obtained.  The
vortex density of the ground state, as a function of the external field, is
a Devil's staircase, with a plateau for every rational value of $\omega$.
The width of each of these steps depends strongly on the approximation made
when calculating the inductance effect: if the self-inductance matrix is
considered, the $\omega=0$ phase tends to occupy  all the diagram as the
penetration depth decreases.
 If, instead,  the whole inductance matrix is
considered, the width of any step tends  to a non-zero value in the
limit of very low penetration depth.
  We have also analyzed the stability of
some simple metastable phases: screening fields are shown to enlarge
their stability range.

\end{abstract}

\vspace{0.9 cm}
\vspace{0.9 cm}

\section{Introduction}

Theoretical research in Josephson junction arrays (JJA) is continuously
progressing through models which involve an increasing complexity. They
represent a better approximation to the understanding and prediction of the
many different interesting phenomena which occur in such
systems.\cite{Frascati} An important contribution to this advance is the
inclusion of current induced magnetic fields (CIMF) developed by different
groups in the last years.\cite{se} 
 Taking CIMF into account is compulsory in order to provide a correct
description of Josephson junction arrays at low temperature, when
the penetration depth of the magnetic field is
of about the cell size.

In all the cases, the study was carried
out through the numerical simulation of the dynamics of the gauge invariant
phase differences. Interest has been mainly focused on the effects of CIMF
in the properties of arrays driven by external currents.  However, we
have no knowledge of studies on the ground state properties of inductive
arrays.

This paper deals with the static properties of a Josephson junction ladder
(JJL),  with anisotropy in the Josephson couplings,
in the presence of an external magnetic field  (figure\ \ref{ladder}). 
In particular, we have 
calculated the ground state phase diagram of a system defined
by a hamiltonian which includes the magnetic energy due to the CIMF, in
addition to the usual Josephson coupling contribution.

Recently two different groups\cite{Mazo,Dennis} have faced the ground state
problem of a JJL in the presence of a magnetic field and in the limit of
infinite penetration depth -- no screening effects --. In this approximation,
some general properties concerning the ground states can be deduced.
In regard to the ground state problem, the hamiltonian describing the system
belongs to a universality class of convex 1-D models of spatially modulated
structures,\cite{Mazo} such as the Frenkel-Kontorova model.\cite{FK}
This fact allows an exact description of the ground states phase diagram and
the relevant elementary excitations of the system.\cite{Mazo}
The diagram, a function of the external field and the anisotropy parameter,
consists of a series of tongues labelled by the vortex density $\omega$.
Both rational and irrational values of $\omega$ are possible, corresponding,
respectively, to commensurate and incommensurate phase configurations.
The vortex density of ground state configurations as a function of the
external magnetic field is a Devil's staircase function, with plateaus for
every rational value of $\omega$. Incommensurate ground states exhibit two
regimes, separated by an Aubry transition \cite{Aubry}: 
below a certain value of the
parameter that describes the anisotropy, the configuration is undefectible
(no defects can be sustained) and unpinned (any external current, though
infinitesimal, causes a non-zero voltage); above this value, the solution
is defectible and pinned.

In this paper we use the work by Mazo et al.\cite{Mazo} as starting point
and include CIMF. We thus obtain a more realistic description of the ground
states and, in general, of the equilibrium properties of the ladder. Such
results may be of interest in understanding experiments in JJL where
relevant parameters can be fixed at will.

We have used different numerical methods -- effective potentials method
combined with root finding methods and stability analysis of solutions, as
well as dynamical relaxation -- in order to find the ground state phase
diagram and other metastable configurations. Results are obtained for three
approximations to the calculation of the induced fields: (A) self-inductance
contributions; (B) self-inductance plus nearest neighbours mutual inductance
contributions; (C) full-inductance matrix. In all the cases the vortex density
of the ground state, as a function of the external field, exhibits a 
Devil's staircase structure.
 Special
attention has been focused on the behaviour of the system in the small
penetration depth limit.  In this limit, a vortex can be described as a
flux quantum concentrated in one only cell.
 The ground state phase
diagram shows an important difference depending on the approximation made
when calculating the inductance effect: if the self-inductance matrix is
considered, the $\omega=0$ phase, in this limit, nearly occupies all the
diagram. When the whole inductance
matrix is considered we find ground states with no null vortex density in a
wide region of the phase diagram.

 The variation of the induced flux with the penetration depth allows 
an estimation of the physically interesting range of values of this
array parameter. We have also studied the dependence of some of the
vortex properties with the penetration depth and the anisotropy of the ladder,
such as its extension and the distribution of gauge-invariant phases and
the induced flux.
Finally, we consider the stability of some simple metastable commensurate
phases when the external field is varied. Notably, the stability
intervals enlarge when the penetration depth decreases, existing a critical
value of the penetration depth for the stability of each phase at zero
external field.

The paper is organized as follows: in section II we introduce the model and
the different methods and approximations used to compute its properties.
Results on the ground state phase diagram and the stability of some simple
commensurate phases are reported in section III. Different approximations to
the calculation of the CIMF are discussed.

\section {Description of the model and the method}

The classical hamiltonian\cite{class} describing the system is:
\begin{equation}
\begin{array}{ll}
H= &  - \sum_{i} \left[
                        J_{x}cos(\theta_{i}-\theta_{i+1}-\pi f_0-\pi f_i)
                        \right. \\
   & \left.
            + J_{x}cos(\theta_{i}^{'}-\theta_{i+1}^{'}+\pi f_0+\pi f_i)
            + J_{y}cos(\theta_{i}-\theta_{i}^{'})
        \right] \\
   &     + {1 \over 2} \Phi_0^2 \sum_{ij}  f_i L^{-1}_{ij} f_j.
\end{array}
\label{hlarges}
\end{equation}
Here $\theta_i$, ($\theta^{'}_{i}$) denotes the phase of the superconducting
order parameter on the upper (lower) branch of the ladder at the $i$th site
(see fig. \ref{ladder}).
$f_0$ is the magnetic flux due to the external field, which is assumed to be
constant along the array. $f_i$ is the induced flux through plaquette $i$, a
function of the currents in the ladder. Both $f_0$ and $f_i$ are expressed
in terms of the flux quantum, $\Phi_0$. Thus, the total magnetic flux
$\Phi^{tot}_i$ through a given plaquette $i$ is
$\Phi^{tot}_i=\Phi^{ext}+\Phi^{ind}_i=\Phi_0(f_0+f_i)$.
The model is periodic in $f_0$ with period 1 and has symmetric
reflection about $f_0=\frac{1}{2}$ in the interval $[0,1]$. Thus we will restrict
our analysis to values of $f_0$ within the interval $[0,\frac{1}{2}]$.

When writing Eq.(\ref{hlarges}) we have made a convenient gauge choice: we
consider that the vector potential is parallel to the long axis of the
ladder, and takes opposite values on the upper and lower branches (see
Fig.\ref{ladder}). In this gauge $f_0$ and $f_i$ are trivially related to
the line integrals of the vector potential $\vec a$ through a link of the
ladder:
$a_{\alpha\beta}={2\pi\over\Phi_0}\int_\alpha^\beta \vec a d\vec l 
= \epsilon\pi(f_0+f_i)$,
where $\epsilon=+1 (-1)$ for upper (lower) links in the ladder and
$\epsilon=0$ for vertical links.

$J_\alpha \; (\alpha=x,y)$, the Josephson coupling energy, is related to the
critical current through the junction, $I_{c\alpha}$, by
$J_\alpha=I_{c\alpha} \Phi_0 / (2\pi)$. The inductance matrix of the
array, $L$, is defined as
\begin{equation}
  L_{ij} = 
{\Phi_0 \over I_{cx}} {1 \over 8 \pi^2 \lambda_\perp} \Lambda_{i j},
\end{equation}
where $\Lambda_{i j}$ is an adimensional matrix containing just geometrical
coefficients (see Appendix A).
$\lambda_\perp$ is the penetration depth, 
defined as in \cite{Orlando}
\begin{equation}
\lambda_\perp = {1 \over {4 \pi^2}} { \Phi_0^2 \over {\mu_0 J_x a}},
\label{lambda_perp}
\end{equation}
being $a$ is the lattice spacing.

It can be seen that the configurations which minimize hamiltonian
(\ref{hlarges}) comply with $\theta_i+\theta^{'}_i=const.$ Fixing this
constant equal to $0$ and normalizing by $J_x$ in order to work with
adimensional quantities we get

\begin{equation}
\begin{array}{lll}
H= &  - \sum_{i} \left[
                        2 cos(\theta_{i}-\theta_{i+1}-\pi f_0-\pi f_i)
                        \right. \\
   & \left.
            + \frac{J_y}{J_x} cos(2\theta_{i})
        \right] \\
   &     + { \Phi_{0}^{2} \over 2 J_x } \sum_{ij} f_i L^{-1}_{ij} f_j,
\end{array}
\label{hsim}
\end{equation}
where the quotient $J_y / J_x$ defines the anisotropy of the ladder.
To solve the ground state phase diagram we will restrict our analysis
to expression (\ref{hsim}).

We consider here three approximations to the inductance matrix: the simplest
model (case A) assumes a diagonal inductance matrix. In this case, the flux
induced in a given plaquette only depends on the mesh current in the same
plaquette. Next step in complexity, case B includes also nearest neighbours
inductances for the couplings between cells. Then, we assume
$L^{-1}_{ij}=\tilde{L}\delta_{ij}+\tilde{M}\delta_{ij\pm1}$. Case C
considers the full-range inductance matrix. In the first case the term in
Eqs.(\ref{hlarges}) and (\ref{hsim}) giving an account of the magnetic
energy becomes
\begin{equation}
H_{magn} = \sum_i d_K f_i^2,
 \;\; {\rm with}  \;\;
 d_K =  8 \pi^3 \lambda_\perp (\Lambda^{-1})_{0 0}. 
\label{Hmagn1}
\end{equation}
In case B,
\begin{equation}
H_{magn} = \sum_i d_K f_i^2 + 
\sum_i \alpha d_K f_i (f_{i-1} + f_{i+1}).
\label{Hmagn2}
\end{equation}

In the system under simulation (we consider
square cells; currents are supposed
to flow  within a cylinder of length $a$ and ratio $0.005 a$)
$d_K  \approx 6.8 \lambda_\perp$ and
$\alpha=\frac{\tilde{M}}{\tilde{L}}\simeq0.21355$.

We have used different methods to solve the problem. In the cases A and B it
is possible to do it using an effective potentials method properly adapted to
study our model,\cite{potefec1} which is numerically equivalent to a 1D
system with 
just next-nearest-neighbours interactions. Effective
potentials method is an efficient method to study the ground state
configurations of such kind of systems in the thermodynamic limit. This
method is based on the computation of certain functions, the effective
potentials, which contain all the relevant information on the relaxation of
local fluctuations to the ground state configurations.\cite{effpot} A long
computation time is required if one wants to obtain the phase diagram of the
system with a high precision. This suggests the convenience of complementing
the method with other procedures.

Effective potentials can provide, within a reasonable amount of time,
approximate solutions to the ground state problem as a function of the
external field, the anisotropy of the ladder and the penetration depth.
Starting from these guesses, one can obtain more precise results by applying
standard root finding methods (calculating stable solutions to
$\frac{\partial H}{\partial x_i}=0$) or even dynamically (letting the
approximate solution relax). 
We make note that root finding methods require
 the use of Eq.\
(\ref{hlarges}) to describe the system since Eq.\ (\ref{hsim}) is just an
adequate expression when dealing with minimum energy configurations, which
is a reduced subspace of the whole system. We have checked that the same
results are obtained if one applies the effective potentials method with a
high precision or if one combines it with any of the complementary
procedures described above. By comparing the energy curves corresponding to
different configurations one can determine the border of the tongues with
different vortex densities.

Moreover, making use of these procedures one can study how a ground state
configuration modifies when the parameters vary. In
this case, it is convenient to keep in mind that, in general, a vortex
configuration is stable beyond the range of parameters in which it is
the ground state solution. There, root finding methods are adequate, and they
must be completed by doing the linear stability analysis of the
solutions. 

Model C involves the total flux matrix. Interactions between the variables
extend to all the lattice and the problem can not be tackled with the
effective potentials method. In this case, we consider the results achieved
in approximation B as guesses and let the system evolve dynamically and 
relax down to the equilibrium. Details about the dynamical algorithm are given
elsewhere.\cite{noi}

\section {Results: ground state phase diagram and stability analysis}

Figure \ref{diagrams}a shows the ground state phase diagram in the case of
infinite penetration depth \cite{Mazo} $\lambda_\perp$ (thus neglecting
CIMF). The different tongues are characterized by the vortex density
$\omega$. This quantity is directly related to the periodicity of the
configuration: a value $\omega=\frac{p}{q}$ implies that 
relevant physical
quantities -- gauge invariant phases differences, induced fluxes... -- are
spatially periodic: every $q$ plaquettes these quantities are exactly
repeated. Here vortices are defined as usual. We make use of the well-known
property of fluxoid quantization to define the vorticity $n_p$ on any
plaquette. The clockwise sum of the gauge invariant phases (restricted to
the interval $( -\pi, \pi]$) along the links of the cell gives $\sum_{\alpha
\beta \in i} (\Theta_\alpha - \Theta_\beta - a_{\alpha\beta}) = 2 \pi (n_p -
f^{tot})$. The vortex density $\omega$ is equal to the spatial average of
$n_p$.

As mentioned in the Introduction, the vortex density as a function of the
external field is a Devil's staircase, with plateaus for every rational
value of $\omega$. Figures \ref{diagrams}b and \ref{diagrams}c show the
phase diagram in the cases A and B, for a penetration depth
$\lambda_\perp=1$, computed using the effective potentials method. Diagrams
(b) and (c) are qualitatively similar to diagram (a). As one expects, the
CIMF tend to push the external magnetic field out of the array: as
$\lambda_\perp$ decreases, the $\omega=0$ tongue grows. 
 There is, however, a remarkable difference between
these diagrams: while in case A all the tongues but the $\omega = 0$ one
compress, in case B the $\omega = \frac{1}{2}$ phase does not shrink, and the rest
of the phases are compressed between these two. In short we will study
carefully the limit of this behaviour when $\lambda_\perp \to 0$.

The effect of the CIMF is to increase the critical field $f_c$ up to which
the $\omega=0$ configuration is the ground state. The devil's staircase is
thus restricted to a narrower interval of values of the field.

The question arises whether CIMF are able to change qualitatively the nature
of the phase diagram. In other words, we want to check if for some
$\lambda_\perp$ the array is able to expulse completely the external field
and, consequently, this tongue occupies all the phase diagram. If this is not
the case, is the devil's staircase structure preserved for all the values of
$\lambda_\perp$?

In order to gain a more complete understanding of the properties of the
model and, in particular, to throw light on the previously raised questions,
it is interesting to study the dependence of the vortex characteristics on
the different physical parameters. Such study is firstly carried out by
considering commensurate phases with a low vortex density (e.g.
$\omega=\frac{1}{128}$) in order to prevent vortex-vortex interaction effects.
In particular, we are interested in studying the vortex extension. It
is directly related to the distribution of the gauge invariant
phases around the vortex barycenter and depends on the values of $J_y$ and
$\lambda_\perp$ (see figure \ref{1v}). 
 In cells
near the vortex centre, the phase decays exponentially. This decay, as
distance increases, becomes  smoother; for large $i$ (where $i$ is 
 the distance to the centre), the phase is of the form
($\phi_i \sim i^{-3}$).
Anisotropy affects the exponential part of the curve: a decrease of
$J_y$ implies a smoother exponential decay, while the long-distance
behaviour remains unchanged. Instead, varying $\lambda_\perp$ makes
the whole curve shift.
 In cells far away enough from the centre, the flux is negative
and its absolute value is a decreasing function of the distance.
The negative flux is due to the sign of the mutual inductance term (see
Appendix A). This feature was already reported by Phillips et al.\cite{se}.
As $\lambda_\perp$ decreases the
vortex becomes more and more localized on its central plaquette and, in the
$\lambda_\perp\to 0$ limit, tends to identify with the fluxoid (the
quantized magnitude defined above, that indicates the barycenter of the
vortex). We can think of $\lambda_\perp$ as the radius of the vortex:
$\lambda_\perp \to 0$ implies that the vortex remains restricted
just to the cell where $n_p = 1$.

When no CIMF are considered ($\lambda_\perp \to \infty$ limit) the total
magnetic flux through a plaquette is just the external flux, which is
constant along all the array. Thus, the flux distribution along the ladder
is independent on the vorticity. Then, there is no flux quantization and the
vortex density is not a flux quanta density but a fluxoid quanta density. On
the contrary, the situation changes drastically when CIMF are taken into
account, being the number and extension of vortices directly connected to
the distribution of the {\em induced} flux along the ladder. Let us consider
the limit $\lambda_\perp \to 0$ and the $\omega = 0$ ground state
configuration; there, the currents tend to uniformly screen the external
field (in every cell $f_i\to - f_0$). Thus, the array exhibits a behaviour
that resembles the Meissner effect: the external field is screened by the
system and no field penetrates the array. For any other value of $\omega$,
the flux distribution is quite different. In the cells where the vorticity
is equal to zero, the induced flux tends to cancel the external one, being
the total flux equal to zero. In the cells where the vorticity is equal to
one, the induced flux $f_i \to (1 - f_0)$, being the total flux equal to one
flux quantum. Then, in the $\lambda_\perp \to 0$ limit the fluxoid
identifies with fluxon, and it is well localized in a cell of the array.
This is illustrated in figure \ref{find}, where the dependence of the
induced flux on $\lambda_\perp$ is shown both in cells with vorticity 0 and 1.
 We have chosen a configuration with $\omega = \frac{1}{2}$, and 
$f_0 = \frac{1}{2}$,
but the behaviour is general: for other values of $\omega $ the fluxes in
cells with zero vorticity depend on the distance to the nearest vortex, but this
difference is of the order of $\lambda_\perp^2$ in the $\lambda_\perp \to 0$
limit. 
We can distinguish 3 regions:
For $\lambda_\perp > 4$, $|f_i| \leq 0.1 f_0$, and considering an
infinite penetration depth is a justified approximation. On the other hand, 
for $\lambda_\perp < 0.12$, we observe the low $\lambda_\perp$ behaviour:
$|f_i| > 0.9 \left( n_i -f_0 \right) $, and
screening fields are dominant. Between them there is an intermediate region, 
around $\lambda_\perp = 0.7$ (where the derivative of the induced flux respect 
to the logarithm of the penetration depth is maximum).

The description presented in the previous paragraph permits to obtain some
simple expressions for the energies of the different configurations for low
values of $\lambda_\perp$. The hamiltonian (\ref{hlarges}) consists of two
components, corresponding to the Josephson and the magnetic energies. We
have numerically checked that, as $\lambda_\perp \to 0$, the first
term saturates first than the magnetic one. Thus, for low enough
$\lambda_\perp$, we can approximate the energy per plaquette by
$E_p=-3+\frac{\Phi_0^2}{2J_xN_p}\;(n_i-f_0+\delta f_i)\;(L^{-1})_{ij}\;
(n_j-f_0 +\delta f_j)$,
where $\delta f_i \sim O(\lambda_\perp)$, $N_p$ is the total number of
cells, and $n_i = 0,1$ is the vorticity of cell $i$.
Let us consider first the case A in this approximation.
The energy per cell of a configuration with $\omega=\frac{p}{q}$ is
$E_p = -3 + d_K \left( \frac{p}{q} \; (1-f_0)^2 + \frac{q-p}{q} \;f_0^2 \right)$.
As $f_0 \in [0, \frac{1}{2}]$, $(1-f_0) \geq f_0$, and $E_p$ is a
increasing function of $\omega$: for any $f_0 \not = \frac{1}{2}$, $\omega_0 <
\omega_1$ implies $E(\omega_0) < E (\omega_1)$ and the configuration with
$\omega = 0$ is the ground state. If $f_0 = \frac{1}{2}$, $E_p$, as
previously defined, has the same value for all $\omega$'s. A second order
approximation in $d_K$ is required. It is easily obtained that
$E(\omega=\frac{1}{2};f_0=\frac{1}{2})<E(\omega=0;f_0=\frac{1}{2})$ (in
particular, $E(\omega=0;f_0=\frac{1}{2})-E(\omega=\frac{1}{2};f_0=\frac{1}{2})=
d_K^2 / (6\pi^2 J_y) $).

We remark that these results correspond to the case A and an extreme limit
($\lambda_\perp\to 0$). Nevertheless, the devil's staircase is observable
down to low values of $\lambda_\perp$ 
 (we have checked this point
even at $\lambda_\perp = 0.012$). As an example, 
 figure \ref{eners} shows the
energy of stable configurations with different values of $\omega$
$(0,\frac{1}{2},\frac{1}{3},\frac{1}{4},\frac{1}{5},\frac{2}{5})$ 
 as a function of the external field when $\lambda_\perp= 0.5$. 
The  energy of the ground state staircase corresponds to the
envelope of the curves,  thus an approximation to the $\omega(f_0)$
function can be obtained from them.

Things change when one considers a more complete approximation to the
inductance matrix (models B and C).  
In this case the $\omega=0$ phase does not
fill the diagram at any value of $\lambda_\perp$ and other commensurate
phases are clearly appreciated. Before performing a rigorous analysis, we
present a plausibility argument in support of this statement. Let's begin by
comparing $\omega = 0$ and $\omega = \frac{1}{2}$ phases. In a $\omega = 0$
configuration, the currents and flux are identical in all the cells. On the
other hand, a configuration with $\omega = \frac{1}{2}$ exhibits a spatial
periodicity with period 2a; when $f_0 = \frac{1}{2}$ the flux and currents on one
cell are of the same module and the opposed sign respect to those on the
adjacent cells. This allows us to define an effective $d_K$ for both
configurations, so that the magnetic energy per cell is just
${d_K}_{eff}f_i^2$ (see Appendix A).
In case B ${d_K}_{eff}(\omega=0) = 10,945\lambda_\perp$ and
${d_K}_{eff}(\omega=\frac{1}{2})= 4,617 \lambda_\perp$
while in case C ${d_K}_{eff}(\omega=0) = 11,176 \lambda_\perp$ and
${d_K}_{eff}(\omega=\frac{1}{2})= 4,638 \lambda_\perp$.
In general, for $w_0 < w_1$ ${d_K}_{eff} (w_0) > {d_K}_{eff} (w_1)$, and thus
$E(w_0; f_0= 1/2) > E(w_1; f_0= 1/2)$. As the energy is a continuous function
of $f_0$ the previous inequality maintains for a range of $f_0$ values
near $f_0 = 1/2$.

Moreover, it is possible to extend in a trivial way the argument previously
developed for case A, and to calculate the energy per plaquette of any
vortex distribution as a function of $f_0$. Note that a configuration with
$\omega=\frac{p}{q}$ is described by a periodic spatial structure 
the basic sub-array of which
consists of $q$ plaquettes containing $p$ vortices. We can thus reduce the
system to one with just $q$ cells. In order to do that it is necessary to
generalize the previous reckoning and to redefine the components of the
matrix $L$ in order to take into account the contribution of each of the
infinite replicas of the basic sub-array.
 Thus the inductance between two cells at
distance $j$ is given by 
$\hat{L}_{j} = \sum_n  L_{0,j+nq} $,
where $n=0,\pm1,\pm2,\ldots$, $j = 0, \ldots, q-1$
and $L_{0,i} = L_{0,-i}$.
In the limit $\lambda_\perp \to 0$ the induced flux in any cell is given by
a vector $F \equiv \{ f_a f_b f_c ...\}$ with $f_i = (1-f_0+\delta f_i)$ or
$(-f_0 + \delta f_i)$ depending on the occupation number of the cell, $n_i$.
Thus, the energy per plaquette, up to the first order in $\lambda_\perp$, is
given by $E_p=-3+\frac{\Phi_0^2}{2J_xN_p}f_i (L^{-1})_{i j} f_j$, with $f_i=
n_i - f_0$. We have computed this expression for a series of values of
$\omega$. By comparing the energies of the curves for different
configurations we have obtained a Devil's staircase, see figure \ref{devil}.
This figure corresponds to the case C; in case B an analogous behaviour is
observed.

Hereafter, we will consider the response of the JJL to continuous
variations of the external field. We will restrict our analysis to the study
of the stability intervals of some simple commensurate phases  which are
the ground state solution at some value of the parameters of the system
(thus, we consider only ordered phases including just vortices (and no
antivortices) and for which $0\leq \omega \leq \frac{1}{2}$).
Such perspective, in the no screening field effects limit, has been studied
in reference \cite{Mazo} in order to characterize the dynamical approach to
equilibrium, which has been shown to lead to slow relaxation. Here we will
just focus on an important difference which appears when CIMF are
considered. Such study is carried out through a quasi-static computation of
ordered stable configurations (local minima of hamiltonian (\ref{hlarges}))
with a determined vortex density, when the external field is slowly varied.

As previously mentioned, the range of stability of any vortex configuration
($\omega$) is broader than the interval of the parameters in which it is the
ground state. In general, there exists a critical value of the field
for the stability of each phase: for $f_0 < f_c (\omega)$ the phase $\omega$
is no more stable.  The loss of stability when $f_0$ decreases
occurs in this way: decrease in the external field implies an increase in
the supercurrent through the horizontal links in order to maintain the
vortex density in the array. The instability of the state takes place when
the supercurrent in one link reaches its maximum value. At this point any
small change in the field can not be sustained by an increase of the
currents and the vortex structure becomes unstable, and the system relaxes
to a new vortex configuration. That would be the process for the changes of
vortex density when the external field is varied or when the thermal noise
is high enough to produce a vortex to jump over the energy barrier of the
metastable phase and then the system approaches to some other more stable
phase.

In the limit of neglecting screening effects \cite{Mazo}
$f_c (\omega) > 0$ for all $\omega \not = 0$, and thus when $f_0=0$ only the
$\omega=0$ phase is stable. 
 The inclusion of CIMF changes this situation.
 As $\lambda_\perp$ decreases the
range of stability of a given phase enlarges.
Moreover, for each configuration $w$ 
there is a
critical value for the penetration depth $\lambda_{\perp c}(w)$:
 if $\lambda_\perp <\lambda_{\perp c}(w)$ 
 the phase
is stable at every value of the external field. 
Let's consider the two extreme cases:  the
configuration containing one single vortex and the $w=1/2$ phase. 
 The repulsive character of the
vortex-vortex interactions implies that the stability of the configuration
containing a single vortex is a necessary condition for the stability of any
phase with $0<\omega \leq \frac{1}{2}$, so that the particular value of
$\lambda_\perp$ at which stability of the configuration occurs
($\lambda_\perp^v (f_0)$) is an upper bound for the stability of the phases 
with
$0<\omega \leq\frac{1}{2}$. On the other hand, stability of the
$\omega=\frac{1}{2}$ phase ensures the stability of any other phase with
$0<\omega<\frac{1}{2}$, so that the particular value of $\lambda_\perp$ at
which stability of the $\omega=\frac{1}{2}$ occurs
($\lambda_\perp^{1/2} (f_0)$) is an lower bound for the stability 
of the
phases with $0<\omega<\frac{1}{2}$. Thus,
$\lambda_{\perp c}^{1/2} (f_0) \leq \lambda_{\perp c}^{\omega_i} (f_0) \leq
\lambda_{\perp c}^{v} (f_0)$.
Figure \ref{estab} shows the regions of stability of the vortex configurations.
For values of the parameters above the curves (region S), any vortex
configuration is stable. In region S-I, as we
move towards the origin of coordinates, the
different states become unstable (in the order of decreasing $w$). 
In I the only stable configuration is that with $w=0$.
Looking again at the supercurrents in the array, we see that as
$\lambda_\perp$ decreases the gauge invariant phase differences of the
metastable configurations approach to zero and thus the supercurrents are
lower, rendering the phase more stable.

\section{Discussion}

In section III we have presented the phase diagram of a Josephson junction
ladder in the presence of screening magnetic field effects. Numerical
results evidence the existence of a series of tongues labelled by the {\it
mean vorticity} $\omega$. Such magnitude exhibits a devil's staircase
structure when the external field is varied.

When CIMF are considered, the system presents a behaviour that
 resembles the Meissner effect: the self-induced field tends to push
the external field out of the array, causing the growing of the $\omega=0$
tongue and the shrinking of the range of parameters where the devil's
staircase is observed. We have compared the results of using different
approximations when calculating the induced fluxes. If only the
self-inductance term is considered, for a value of $\lambda_\perp$ low
enough the $\omega=0$ phase occupies all the phase diagram except for a tiny
region near the $f_0=\frac{1}{2}$ line. However states with inserted 
fluxons are
stable and, moreover, their range of stability increases as $\lambda_\perp$
decreases. 
Instead, if one takes into account the whole
inductance matrix, the critical field for the $\omega=0$ phase (the
frustration above which the configuration is no more the ground state)
remains lower than $\frac{1}{2}$ for all the values of the penetration depth. 
Thus  commensurate phases with vortices are always clearly visible  in the
phase diagram. 

 The three approximations made: 
A (self-inductance), B (self-inductance plus nearest neighbours mutual 
inductance terms) and C (full-inductance matrix) correspond to different
distributions of the relative weights of the inductance matrix components.
We remark that these distributions are a function of the
geometry of the currents flowing in the array (see Appendix).  
Let's consider the case in which currents flow inside cylindric tubes of
radius $r$ and length equal to the lattice spacing $a$ (the qualitative
conclusions can be extended to any kind of cross section).
If $r << a$, 
$\Lambda_{i,i} > |\Lambda_{i,i+1}| >> |\Lambda_{i,i+j}|(j>1)$ and approximation
B (considering just the self-inductance plus the nearest neighbours terms
in $\Lambda$) is justified. As $r$ increases, the terms 
 $\Lambda_{i,i+j} (j>1)$ also increase, but are yet too small. They give just
small corrections to the final results.
 In a narrow range of $r$ values around $r \sim 0.25$ the 
dominant contribution to the inductance is self-term (and A is a good
$0-th$ order approximation). Finally, for greater $r$,
$  |\Lambda_{i,i+j}| / \Lambda_{i,i} $ cannot be neglected and considering
the whole inductance matrix is compulsory.

This behaviour (the existence of an infinite set of ground states as
the parameters vary which show a Devil's staircase structure) is
characteristic of a broad class of spatially modulated structures with
convex interparticle interactions. In the limit of neglecting
screening field effects ($\lambda_\perp\to\infty$) it is well
established the equivalence, regarding the ground state problem, of
hamiltonian (\ref{hlarges}) with a one-dimensional XY model with anisotropy
and the ground state problem of the system is equivalent to the one of a
Frenkel-Kontorova model with convex interparticle interaction, which allows
for applying the Aubry theory for this class of models. 

However, and despite the qualitatively similar behaviour shown by our
simulations, the inclusion of CIMF renders it difficult to establish
any equivalence between model (\ref{hlarges}) and the 1-D models named
above. This point is  beyond the scope of this paper and remains as an open
question deserving future research.

The introduction of the CMIF allows to study the continuous variation of the
system from the full penetration of the external field
($\lambda_\perp\to\infty$; $\Phi^{tot}_i=f_0$, for $\lambda_\perp \geq 4$) 
limit to the case of an array
where all magnetic flux, in the first order of $\lambda_\perp$,
 appears quantized ($\Phi^{tot}_i= 0$ or $1 + O(\lambda_\perp)$), 
for $\lambda_\perp \leq 0.12$). This
transition is reflected in figure \ref{find}. 

An appealing question is that of the behaviour -- analysis of the robustness
and stability specially -- of different metastable configurations of a
system under the variations of the external parameters. It provides
information on the energy landscape of a system and other properties of the
phase space which can determine interesting situations such as a constrained
dynamics or slow relaxation processes. The inclusion of the CIMF enlarges
the stability range of the vortex configurations, as we have explicitly
shown in the extreme cases of one single vortex and a $w= \frac {1}{2}$ phase.
Quite recently Hwang, Ryu and Stroud \cite{HRS} have extended the results on
the ground state properties of the Josephson junction ladders in the large
penetration depth limit, to treat the IV characteristics of a ladder array.
They found that when "current is injected perpendicular to the ladder edges,
the critical current is unchanged from its $f=0$ value up to a penetration
field of $f_{c1}\simeq0.12$ flux quanta per plaquette". 
This result can be easily understood from the considerations on the stability
of the one vortex configuration we have made above. 
At low values of the external field the ground state phase configuration of a
bidimensional square Josephson junction array has a vortex density different
from zero and the existence of vortices produces a critical current lower than
that of the $\omega=0$ phase.
In the ladder, however, the situation is quite different:
at low values of $f$ the ground state in the ladder is the $\omega=0$ phase
and, moreover, as we have described above, the
configuration with just one vortex is unstable and the $\omega=0$ phase is
the only stable attractor for arbitrary initial phase configurations.
Consequently, at low values of the external field the critical current of the
ladder is expected not to change, since it depends on the vortex density.
However, we have shown that the critical value of the field for the stability
of the one single vortex configuration is $f_c^{v}=0.1175\pm0.00065$ for an
isotropic ladder. 
For values of the field above $f_c^{v}$ different vortex
configurations are stable; thus, arbitrary initial phase configurations
generically relax to different possible metastable configurations, which can
be described as irregular arrays of vortices. In that situation the critical
current is essentially associated with the depinning transition of
these vortex phases, which occurs at a lower value of the external current.
Hwang, Ryu and Stroud give a value around $0.12$, quite close to our
prediction of $f_c^{v}=0.1175\pm0.00065$.

As we have seen above, as $\lambda_\perp$ decreases, the value of
$f_c^{v}$ also decreases, vanishing at $\lambda_\perp=1.812\pm0.018$. 
Then, the diagram of stability (Fig. \ref{estab}) allows to make the following
conjecture on the behaviour of the critical current of the Josephson junction
ladder when current induced magnetic fields are taken into account:
as the penetration depth decreases, the range of values of the external field
for which the critical current remains unchanged \cite{HRS}, also decreases
and it vanishes when $\lambda_\perp=1.812\pm0.018$. This conjecture is based
on the natural association between the critical current of the
ladder and the stability of the one vortex configuration.

\section*{Acknowledgements}

We are indebted to C. Giovannella, F. Falo and L. M. Flor\'{\i}a for many
useful discussions on this and related subjects. JJM thanks C. Giovannella,
J. C. Ciria and the Dipartimento di Fisica in the Universit\`{a} di Roma Tor
Vergata for their hospitality and making their facilities available to him.
JCC is supported by a post-doctoral grant from MEC (Spain). JJM is supported
by a grant from from MEC (Spain), the Project No. PB92-0361 (DGICYT) and
European Union (NETWORK on Nonlinear Approach to Coherent and Fluctuating
Processes in Cond. Matt. and Opt. Phys.; ERBCHRXCT930331).

\section*{Appendix A: the inductance matrix}

The $L$ matrix is obtained in the following form: we have applied the
Biot-Savart law in order to calculate the magnetic field induced on a link
by all the currents circulating in the array. We thus obtain
\begin{equation}
 \int_\alpha^\beta \vec a^{ind} d \vec l = 
 \sum_{\gamma \delta} 
FF_{\alpha \beta; \gamma \delta} I_{\gamma \delta} , 
\label{aij_dim}
\end{equation}
where $FF$ is the form matrix,  which depends on the geometry of the
array, and $I_{\alpha \beta}$ the total current passing
through link $\alpha \beta$. If the links $\alpha \beta$ and 
$\gamma \delta$ are perpendicular we take $FF_{\alpha \beta; \gamma \delta} = 0$.

 The self-inductance term 
$FF_{\alpha \beta;\alpha \beta}$ depends strongly on the form of the current.
If, for example, it is supposed to flow within a tube of length $l$ 
with circular cross
section of radius $r$ it is given by
\begin{equation}
FF_{\alpha \beta;\alpha \beta} = 2 l \left( log \left( \frac{2 l}{r} \right)
- \frac {3}{4} \right). 
\end{equation}
If $l$ is measured in meters, $FF$ is given in  $10^{-1}$ Henries.

Instead, the mutual inductances between different links are sensibly the same 
as that
of the filaments through the centers of their cross sections,  even when the
links are very close. In particular, in the case of cylindric currents, the 
mutual inductances are absolutely independent from the radius.

In order to express \ref{aij_dim} as a function of adimensional quantities
we introduce $ff  = \frac {4 \pi} {\mu_0 a}   FF$, thus obtaining
\begin{equation}
 a_{\alpha \beta}^{ind} = {2 \pi \over \Phi_0} 
\int_\alpha^\beta \vec a^{ind} d \vec l = 
{1 \over {4 \pi \lambda_\perp }} \sum_{\gamma \delta} 
ff_{\alpha \beta; \gamma \delta} i_{\gamma \delta} , 
\label{aij}
\end{equation}
where currents are  normalized to the critical current of the link
$I_c$ (in the case of an anisotropic ladder, where $I_x \not = I_y$, we take
$I_c$ as $I_{cx}$). $\lambda_\perp$ is the penetration depth,
see eq. (\ref{lambda_perp}). 

It is easy to obtain an equivalent description of the self-field effect
in terms of the mesh currents and the magnetic flux, as required in equation
(\ref{hsim}). The magnetic flux on $i$-cell is given by
\begin{equation}
\Phi_i^{ind} = \oint_i \vec {a}^{ind} d\vec {l} =
{\Phi_0 \over 2 \pi}\sum_{\alpha \beta \in i } a_{\alpha \beta}^{ind},
\label{p1}
\end{equation}
where the sum is over the four links $\alpha \beta$ of cell $i$, and can be 
expressed
by a linear equation of the form 
\begin{equation}
\sum_{\alpha \beta \in i} a_{\alpha \beta}^{ind} = 
A_{i; \alpha \beta} a_{\alpha \beta}^{ind}. 
\label{p2}
\end{equation}
We have used greek and roman symbols to
denote, respectively, links and cells.

On the other hand, also the mesh currents ($i_i$) are related to the 
link currents ($i_{\alpha \beta}$)
through a linear operator
\begin{equation}
i_{\alpha \beta} = \sum_{\alpha \beta \in i} B_{\alpha \beta;i} i_{i}.
\label{p3}
\end{equation}

Combining (\ref{aij}), (\ref{p1}), (\ref{p2}) and (\ref{p3}) we obtain
\begin{equation}
\Phi^{ind}_i = L_{i j} I_j \; , \;  L_{ij} = 
{\Phi_0 \over I_c} {1 \over 8 \pi^2 \lambda_\perp} \Lambda_{i j} \; , \;
\Lambda_{i ; j} = \sum_{\alpha \beta} \sum_{\gamma \delta}
A_{ i ; \alpha \beta} FF_{ \alpha \beta ; \gamma \delta} 
B_{\gamma \delta; j }.
\label {Lambda}
\end{equation}
  The $\Lambda_{i;j}$ elements depend on the distance $|i-j|$ between the
cells considered. The general properties of matrix $\Lambda_{i ; j}$ are:
$\Lambda_{0 0}$ is positive, and $\Lambda_{i ; j} < 0$ for $ i \not = j$;
for two cells far away enough
($| i - j|  \geq 10$)
the mutual inductance
is $|\Lambda_{i j }|  \sim |i - j|^{-3}$,
as calculated in \cite{J&D}.

  In the following, we will consider that currents flow inside cylindric
tubes with circular cross section of radius $r$.
  If currents are supposed to have a very small cross section (compared to
the lattice spacing, $a$), 
then $ff_{\alpha \beta; \alpha \beta} >> ff_{\alpha \beta; \gamma \delta}$.
Then it can be easily verified that $\Lambda_{i i} \sim 8 \log (1/r)$, 
and $\Lambda_{i, i+1} \sim - 2 \log (1/r)$. The other terms 
$|\Lambda _{i,i+j}|_{(j>1)} << \Lambda_{ii}$ and can be neglected.
As the cross section increases, 
 $ff_{\alpha \beta; \alpha \beta}$ becomes smaller while the rest of the 
$ff's$ remain sensibly the same; $\Lambda_{i,i}$ decreases, 
$\Lambda_{i,i+1}$ increases and changes sign at $r \sim 0.24 a$ and 
$\Lambda_{i,i+j}, (j>1)$ remain the same. 
There is a range of $r$ values $r \in [0.2, 0.28]$ where
 $| \Lambda_{i,i} | > 20 | \Lambda_{i,j} | (j \not = i)$.

  Thus we can establish $r$ ranges where the different approximations A,B and C
made in the paper are acceptable. If $r << a$, 
$\Lambda_{i,i} > |\Lambda_{i,i+1}| >> |\Lambda_{i,i+j}|(j>1)$ and approximation
B (considering just the self-inductance plus the nearest neighbours terms
in $\Lambda$) is justified. As $r$ increases, the terms 
 $\Lambda_{i,i+j} (j>1)$ also increase, but are yet too small. They give just
small corrections to the final results.
 In a narrow range of $r$ values around $r \sim 0.25$ the 
dominant contribution to the inductance is self-term (and A is a good
$0-th$ order approximation). Finally, for greater $r$
$  |\Lambda_{i,i+j}| / \Lambda_{i,i} $ cannot be neglected and considering
the whole inductance matrix is compulsory.

In our calculations, we have considered square cells; for the study of case C
we have considered an intermediate value of $r$ (currents are supposed
to flow  within a cylinder of length $a$ and ratio $0.005 a$ \cite{Nuvoli}).
In this case,
$\Lambda_{0 ; 0} = 38.194$ and
$\frac{\Lambda_{0 ; i}}{\Lambda_{0 0}} \equiv \{$
1, -0.20332, -0.0040118, -0.0010570 ... $ \}$.

We can now calculate the equivalent $d_K$ in the case $f_0 = \frac{1}{2}$ for 
the
configurations $\omega = 0$ and $\omega = \frac{1}{2}$. This can be easily 
made if
one considers the spatial periodicity of the solutions. If $\omega = 0$ all
the cells in the array have the same flux and current; thus we can define an
effective self-inductance matrix as
\begin{equation}
f_i = \left( L_{i;i} + 2*(L_{i;i+1} + L_{i;i+2} + ...) \right) i_i  =
L_{eff} i_i, 
\end{equation}
where factor 2 is due to the sum of the contributions from the cells on the
left and on the right. The terms $L_{i j}, i \not = j$ are negative and
thus $L_{eff} < L_{i i}$. Now ${d_K}_{eff}(\omega=0) =
\frac{8\pi^3}{L_{eff}} = 11,176 \lambda_\perp$.

In an analogous way, one can calculate the value of $L_{eff}$ for a $\omega
= \frac{1}{2}$ solution: for $f_0=\frac{1}{2}$ the flux and the current in one cell have the
same magnitude and the inverse sign of those in the adjacent plaquettes.
Thus
\begin{equation}
f_i = \left( L_{i;i} + 2*( - L_{i;i+1} + L_{i;i+2} - ...) \right) i_i  =
L_{eff} i_i, 
\end{equation}
where now $L_{eff} > L_{i i}$ and ${d_K}_{eff}(\omega=\frac{1}{2}) =
8\pi^3 / L_{eff} = 4,638 \lambda_\perp$.

\begin{figure}
\caption[]{Schematic representation of the Josephson junction array we study
here: an anisotropic ($J_x \neq J_y$) ladder, in the presence of an external
field. The sites denote superconducting islands and the crosses the junctions
themselves. Right-most plaquette shows the mesh current $i_i$ and
the gauge choice, here
$f_i^{tot}=f_0+f_i$ where $f_0=\frac{Ha^2}{\Phi_0}$ is the flux due to the
external field and $f_i$ the induced flux in the plaquette $i$.}
\label{ladder}
\end{figure}

\begin{figure}
\caption[]{Ground state phase diagrams of the JJL obtained using the method
of effective potentials. Each phase is defined by the value of $\omega$ and,
for clarity, only a few of the transition lines are represented.
Figure (a) shows the results for the no screening field case
(or $\lambda_\perp \to \infty$).
(b) Phase diagram for a $\lambda_\perp=1.0$
ladder using approximation A (diagonal inductance matrix) to the calculation
of the induced fluxes.
(c) Phase diagram for a $\lambda_\perp=1.0$
ladder using approximation B (self plus nearest neighbours inductances)
to the calculation of the induced fluxes.}
\label{diagrams}
\end{figure}

\begin{figure}
\caption[]{ 
Vortex shape as a function of the penetration depth $\lambda_\perp$
and the anisotropy. We consider an isotropic
 128-cell ladder
with just one vortex. The whole inductance matrix is used. $f_0 = 0$.
The shape is symmetric, and we just draw the 
right-half of the vortex
(the origin of coordinates is the central
plaquette of the ladder).

The figure  shows the gauge-invariant phase difference along the
horizontal links belonging to the upper branch of the ladder. 
 We compare the cases with 
$J_y = J_x$, $\lambda_\perp = 1$ (black circles)
$J_y = 0.4 J_x$, $\lambda_\perp = 1$ (squares)  and
$J_y = J_x$, $\lambda_\perp = 0.012$ (rhombs).

The inset shows the  induced flux around the central plaquette, in the same 
cases as before.

}
\label{1v}
\end{figure}

\begin{figure}
\caption[]{ Induced flux as a function of $\lambda_\perp$ for a configuration
$\omega = \frac{1}{2}$, at $f_0 = \frac{1}{2}$, $J_x = J_y$.
We draw the flux through two plaquettes with vorticity 0 and 1, respectively.
We consider the C case: whole inductance matrix.
In the inset we show the derivative of the induced flux respect to the logarithm
of the penetration depth. We can distinguish three regions: the extreme limits
$\lambda_\perp > 4$ (the induced flux is $f_i = 0 + \delta / \lambda_\perp$)
and $\lambda_\perp < 0.12 $ ( $f_i = n_i - f_0 + \delta \lambda_\perp$)
and an intermediate region around $\lambda_\perp \sim 0.7 $, where the 
derivative is maximum.
}
\label{find}
\end{figure}

\begin{figure}
\caption[]{Energies of different configurations
as a function of frustration ($\epsilon_\omega (f_0)$) for a penetration depth
$\lambda_\perp = 0.5$ in the case A. We study configurations
$w$ = 0, 1/5, 1/4, 1/3, 2/5 and 1/2 (in order of decreasing slope).
We consider an isotropic ladder ($J_x = J_y$).
The ground state energy at each value of $f_0$ is given by
the envelope of the curves. The inset shows the value of $w$
corresponding to the minimum energy curve for each 	 $f_0$.
}
\label{eners}
\end{figure}

\begin{figure}
\caption[]{Devil's staircase observed in case C in the
limit $\lambda_\perp \to 0$ for an isotropic ladder, calculated as explained
in the text.
}
\label{devil}
\end{figure}

\begin{figure}
\caption[]{ Border between stability and instability regions in the parameter
space for the extreme cases of one single vortex (black points) and $w = 1/2$
configuration (rhombs) in an isotropic ladder. 
In the first case we have considered a 128-cell ladder.

We have checked that  the curves fit to a quadratic function:
$f_0 = \alpha \left( \beta + \frac{1}{\lambda_{\perp c}(f_0)} \right) *
\left(\frac{1}{\lambda_{\perp c}(0)} -   
\frac{1}{\lambda_{\perp c}(f_0)} \right)$.
$\lambda_{\perp c}(0)$ is the penetration depth below which a configuration
is stable at $f_0 = 0$, and 
$\alpha \beta  \frac{1}{\lambda_{\perp c}(0)} = f_c$, where $f_c$ is the
value of the frustration below which a configuration is no more stable
in the limit of no inductance ($\lambda_{\perp} \to \infty$).
For a configuration with one single vortex, 
$f_c = 0.1175 \pm 0.00065$ and $\lambda_{\perp_c}(0) = 1.812 \pm 0.018$;
in the $w=1/2$ case, 
$f_c = 0.215 \pm 0.001$ and $\lambda_{\perp_c}(0) = 1.197 \pm 0.006$.

}
\label{estab}
\end{figure}

\end{document}